\documentclass{PoS}

\usepackage{amssymb}
\usepackage{fontenc}
\usepackage{times}
\usepackage{mathptmx}
\usepackage{graphicx}

\title{Timelike structure functions and hadron multiplicities}

\ShortTitle{Timelike structure functions and hadron multiplicities}

\author{\speaker{Paolo Bolzoni}
\\      
        II. Institut f\"ur Theoretische Physik, Universit\"at Hamburg,\\
	Luruper Chaussee 149, 22761 Hamburg, Germany\\
        E-mail: \email{Paolo.Bolzoni@desy.de}}

\abstract{In this talk we discuss the results obtained in the new approach that we recently introduced to 
consider and include both the 
perturbative and nonperturbative contributions to the evolution of the gluon and quark avarage multiplicities. 
We report on our progresses in solving a longstanding puzzle of QCD. The new formalism is motivated by
recent important theoretical developments in timelike small-x resummation which are also discussed
mostly from an historical point of view. 
We have extended our global
analysis to fit the available data adding the strong coupling constant as a fit parameter. In this way  
our best fit gives $\alpha_s(M_z)=0.124\pm 0.005$
and for the corresponding $\chi^2$ we have obtained a further improvement.}
 
\FullConference{XXI International Baldin Seminar on High Energy Physics Problems,\\
		September 10-15, 2012\\
		JINR, Dubna, Russia}

\begin{document}

\section{Introduction}

When jets are produced at colliders, they can be initiated either by a gluon or by a quark.
Because of the fact that these two partons carry different color charges and spin, it is 
expected that the corresponding jets show different properties: 
typically a gluon jet is braoder and contains a larger amount of hadrons.
Jets with different parent partons can also be distinguished and investigated looking for the jet charge distribution
as discussed in detail in Ref.\cite{Waalewijn:2012sv} with important applications for the LHC.
The interactions between
quarks and gluons are described by quantum chromodynamics (QCD) and to understand the
difference between a quark and a gluon jet is one of the strongest 
test of this theory. 

The typical way to depict the production of a jet from a parton is the following:
an initial quark (or gluon) start radiating gluons, which in turn can radiate 
further gluons or split into secondary quark-antiquark pairs. In this way the virtuality
of the parent parton decreases in a so called parton showering process. Finally when the 
virtuality falls below a certain cutoff the cascade stops and the final state partons 
hadronize into color neutral hadrons, a process usually described by phenomenological models.
This happens because the production of hadrons is a typical process where  
non-perturbative phenomena are
involved. However, for particular observables, this problem can be avoided.
In particular the ``counting'' of hadrons in a jet which is initiated at a certain
scale $Q^2$ belongs to that class of observables and in this case one  
can assume with quite high accuracy the hypothesis of Local Parton-Hadron Duality (LPHD) 
which simply states that parton distributions are just renormalized in the hadronization
process without changing their shape \cite{Azimov:1984np}.
This would in principle allow perturbative QCD at fixed order to make predictions
without the need to consider phenomenological models of hadronization. Nevertheless  
these are processes dominated by soft gluon emissions and it is a well known fact that
in such kinematic regions of the phase space fixed order perturbation theory fails and
resummation is needed (see \emph{e.g.} Ref.\cite{Mueller:1981ex}) and this is the main 
topic discussed in this talk. 

The gluon and quark multiplicities
$\langle n_h(Q^2)\rangle_{g}$ and $\langle n_h(Q^2)\rangle_{s}$ represent
the avarage number of hadrons in a jet initiated by a gluon or a quark respectively at scale $Q^2$.
In the past analytic predictions have been achieved solving the equations for the generating 
functionals in the modified 
leading logarithmic approximation (MLLA)  
in Ref.\cite{Capella:1999ms} up to the so called N$^3$LOr in the expansion parameter 
$\sqrt{\alpha_s}$ \,\emph{i.e.} $\alpha_s^{3/2}$.
However for the ratio  $r=\langle n_h\rangle_g/\langle n_h\rangle_s$
the theoretical prediction is about 10\% higher than the data at the
scale of the $Z^0$ vector boson and the difference with the data becomes 
even larger at lower scales even if the convergence of the perturbative series 
behaves very well. 
An alternative approach was given in Ref.
\cite{Eden:1998ig} where equations for the derivative of the ratio of the
multiplicities are obtained in the MLLA 
within the framework of the colour dipole model. 
There a constant of integration 
which is supposed to encode non-perturbative contributions is fixed by
the data. A constant offset to the quark and gluon multiplicities
has also been introduced in \cite{Abreu:1999rs}. 

Very recently a new formalism has been proposed and developed in Refs.\cite{Bolzoni:2012ed,Bolzoni:2012ii},
where the problem of the apparent good convergence of the perturbative series is solved and where
any \emph{ad hoc} offset is needed once one includes the effects coming
from the full mixing between quarks and gluon evolutions. The result looks like a generalization
of the result obtained in Ref.\cite{Capella:1999ms}.
In the new approach the 
non-perturbative physics to the gluon-quark multiplicities is parametrized 
simply in the initial conditions of the evolution equations. 
Due to the good agreement with the data obtained in Ref.\cite{Bolzoni:2012ii}, 
at the time of the contribution to this conference we have done a study where we have extended our analysis  
adding as a fit parameter also the strong coupling constant.
Thanks to the new outstanding results in small-x timelike resummation obtained in \cite{Kom:2012hd} 
in the $\overline{MS}$ scheme,
we were able to compute exact next-to-next-to-leading-logarithm (NNLL) contributions to the evolution of 
the multiplicity with
with approximated NNNLO normalization factors in the $\sqrt{\alpha_s}$ expansion. Previously published results where 
available up to the NLL accuracy but in a massive gluon scheme which in general is unfortunately
not suitable to combine resummation with fixed higher order corrections which are naturally
given in the $\overline{MS}$ scheme. See Refs.\cite{Albino:2011bf,Albino:2011si} for a 
general discussion about the scheme choice and scheme dependence in this context.

In the following after reviewing the history which lead
to the latest important improvements in ``timelike'' QCD of Refs.\cite{Almasy:2011eq,Kom:2012hd} 
and sketching our formalism, we present, as anticipated, the result of a slightly different global fit than in
\cite{Bolzoni:2012ii} where
also the strong coupling contstant has been added among the free parameters confirming the good quality of 
the fit.

\section{Exiting times for ``timelike'' QCD}

The important and fundamental results obtained in Refs.\cite{Almasy:2011eq,Kom:2012hd} represent the 
beginning of the
happy end of a quite long story: In the 1972 it was realized in Ref.\cite{Gribov:1972ri} that at the lowest order 
the ``timelike'' 
splitting functions (occuring in Semi-Inclusive electron positron Annihilation (SIA)) 
are the same as their ``spacelike'' counterparts (occuring in Deep Inelastic Scattering (DIS)). 
This property goes under the name of Gribov-Lipatov relation. Then in the 1980 it has been 
shown  by Curci, Furmanski and Petronzio \cite{Curci:1980uw} that the Gribov-Lipatov relation
is violated. In that period several groups obtained the NLO contributions to the `timelike`
splitting functions 
\cite{Curci:1980uw,Furmanski:1980cm,Kalinowski:1980wea,Kalinowski:1980ju,Floratos:1981hs,Munehisa:1981ke}. 
In 2004 the NNLO ``spacelike'' splitting functions were published by Moch, Vermaseren and Vogt 
in Ref.\cite{Moch:2004pa,Vogt:2004mw}.
In this last case the calculation could be performed via forward scattering amplitudes, a fact that has
allowed a direct calculation in terms of Feynman diagrams. This is however not the case for the 
``timelike'' splitting functions and different techniques have been investigated. 
Two of them have been the most successfull. The first one is due to Dokshitzer, Marchesini and Salam 
\cite{Dokshitzer:2005bf}
who developed a formalism trying to rescue (at least at the formal level) the Gribov-Lipatov relation
at higher orders. The approach of Ref.\cite{Dokshitzer:2005bf} shaded light on many theoretical aspects
revealing (using their words) ``intrinsic beauty of the perturbative quark-gluon
dynamics''\cite{Dokshitzer:2008zz}. The second one is based on the fact that the ``timelike'' and the 
``spacelike'' splitting functions can be related by the analytic continuation of the scaling variable
$x\rightarrow 1/x$ with $x$ representing the fraction of the parton longitudinal momentum. 
In particular, even if beyond the LO this cannot be done directly from the splitting
functions, it can be done for the corresponding physical evolution kernels and this has been shown in Refs.
\cite{Blumlein:2000wh,Stratmann:1996hn}.  
Formally, the evolution of a structure function is governed by a physical kernel $K$ which depends on the
coefficient function $C$ and the splitting function $P$ according to:
\begin{equation}
Q^2\frac{\partial F}{\partial Q^2}=K\otimes F=\left[\left(\beta(\alpha_s)\frac{dC}{d\alpha_s}+
C\otimes P\right)\otimes C^{-1}\right]\otimes
F,\label{physkern}
\end{equation}  
where $\beta$ is the QCD beta function for the running of the strong coupling $\alpha_s$
\begin{equation}
\beta(\alpha_s)=Q^2\frac{\partial \alpha_s(Q^2)}{\partial Q^2},
\end{equation}
and where $\otimes$ means the usual integral convolution with respect to $x$.
Eq.(\ref{physkern}) also tells us that
\begin{equation}
F=C\otimes D,\label{factor}
\end{equation}
where $D$ is what we call fragmentation function (the analog of the parton densities in the ``spacelike'' case)
and that
\begin{equation}
Q^2\frac{\partial D}{\partial Q^2}=P\otimes D.\label{ap0}
\end{equation}
According to Eq.(\ref{physkern}) and the discussion above it, we see that in principle all the ``timelike'' splitting functions can be obtained 
from the ``timelike'' physical kernels (obtained by the analytic continuation $x\rightarrow 1/x$), 
once the ``timelike'' coefficient functions $C$ are known.
The ``timelike'' coefficients functions are known at NNLO
\cite{Rijken:1996npa,Zijlstra:1991qc,vanNeerven:1991nn,Rijken:1996ns,Moch:1999eb,Blumlein:2006rr,Mitov:2006wy},
while the analytic continuation of the physical kernels from the NNLO ``spacelike'' ones has been obtained in
Refs.\cite{Mitov:2006ic,Moch:2007tx} and finally in Ref.\cite{Almasy:2011eq}. Here we mention only that the 
analytic continuation $x\rightarrow 1/x$ becomes subdle (see Ref.\cite{Stratmann:1996hn}) 
for the logarithmic terms singular in $x\rightarrow 1$
for which
\begin{equation}
\ln (1-x)\rightarrow \ln (1-x) -\ln x +i\pi,
\end{equation}
and that additional constaints (see Ref.\cite{Almasy:2011eq} and referencies therein) 
from the momentum sum rules, the supersymmetric limit ($C_A=C_F=n_f$)
and from the relations found in Ref.\cite{Dokshitzer:2005bf} are needed to close the problem
in a satisfactory way. 

Nevertheless the NNLO ``timelike'' splitting functions for the quark-singlet system obtained in 
Ref.\cite{Almasy:2011eq} are not the end of the story. Indeed they present singularities
in the threshold ($x\rightarrow 1$) and in the large energy limit ($x\rightarrow 0$).
Both kind of singularities are due to the radiation of soft gluons and they make perturbation theory to fail.
As already mentioned, the difficulty with these singularities are overcome by resummation. For very recent
developments in the large-x resummation see \emph{e.g.} 
Refs.\cite{LoPresti:2012rg,Grunberg:2011gx}.
Here we are mainly interested in the small-x resummation because it is directly related to the 
computation of hadronic multiplicities as already realized a long time ago \cite{Mueller:1981ex}.  
With respect to the multiplicity studies, the basic equation is the one governing the evolution of 
fragmentation function $D$ for the gluon-singlet system Eq.(\ref{ap0}), which
in Mellin space sees the convolutional product $\otimes$ turn into an ordinary matrix product:  
\begin{equation}
\label{ap}
Q^2\frac{\partial}{\partial Q^2} \left(\begin{array}{l} D_s(\omega,Q^2) \\ D_g(\omega,Q^2)
\end{array}\right)=\left(\begin{array}{ll} P_{qq}(\omega,\alpha_s) & P_{gq}(\omega,\alpha_s) 
\\ P_{qg}(\omega,\alpha_s) & P_{gg}(\omega,\alpha_s)\end{array}\right)\left(\begin{array}{l} D_s(\omega,Q^2) \\ D_g(\omega,Q^2)
\end{array}\right).
\end{equation}
Here $\omega=N-1$ with $N$ the standard Mellin moments
with respect to $x$ and where $P_{ij}$ are splitting functions as introduced in Eq.(\ref{physkern}). The standard definition for the hadron multiplicity in terms of the 
fragmentation function is given by (see \emph{e.g.} Ref.\cite{Ellis:1991qj}) the 
integral over $x$ of the fragmentation function which is just the first Mellin moment ($\omega=0$):
\begin{equation}\label{multdef}
\langle n_h(Q^2)\rangle_{a}\equiv \left[\int_{o}^1 dx \,x^\omega\,D_a(x,Q^2)\right]_{\omega=0}=D_a(\omega=0,Q^2),
\end{equation}
where $a=s$ for a quark jet and $a=g$ for a gluon jet.
It is clear from the definition in Eq.(\ref{multdef}) that eventual non-integrable singularities
in $x\sim 0$ in the inegrand could not occur for the definition to make sense. We remind that
the singularities in $x=0$ are translated into singularities in $\omega=0$. We have already mentioned
that resummation which includes the singularities from all orders according to a certain
logarithmic accuracy is the standard way to solve this problem. Indeed after resummation, the singular
behavior in $x\sim 0$ (or equivalently in $\omega\sim 0$) disappears. The generally better choice of the
$\overline{MS}$ scheme for to perform resummation has been considered and solved only quite
recently. Firstly the NLL accuracy has been obtained in Refs.\cite{Vogt:2011jv,Albino:2011cm}
and finally the NNLL accuracy has been reached in Ref.\cite{Kom:2012hd}. Here thanks to the
approach used by A.\,Vogt in Ref.\cite{Vogt:2011jv}, the singularities of the splitting functions
are iteratively extracted according to the all order factorization into the transition function $Z(\epsilon)$ 
of the $\epsilon=0$
poles in dimensional regularization ($d=4-2\epsilon$). 
Indeed, according to the factorization theorem \cite{Ellis:1978ty,Curci:1980uw}, we can rewrite Eq.\ref{factor}
in Mellin space as
\begin{equation}
F=C\,D=C^0(\epsilon)Z^{-1}(\epsilon)\,Z(\epsilon)D^0,\label{trans}
\end{equation} 
where $C^0$ and $D^0$ are the ``bare'' coefficient function and fragmentation function.
$Z(\epsilon)$ is the transition function containing only poles in $\epsilon=0$ that are factored out
from $C^0$.
Hence, substituting Eq.(\ref{trans}) into Eq.(\ref{ap0}) we get that
the splitting functions can be directly related to the transition function in the following way:
\begin{equation}
P=Q^2\frac{\partial Z}{\partial Q^2} Z^{-1}=\beta(\alpha_s)\frac{\partial Z}{\partial \alpha_s}Z^{-1}.
\end{equation}
Now it is shown in Ref.\cite{Vogt:2011jv} how one can solve this equation in $Z$ obtaining
at all orders the three highest order poles in $\epsilon$ knowing the NNLO corrections to $P$ and $\beta$. 
Additionally knowing the higher order corrections to $C$ which is pole free one 
obtains from the NNLO computations the all order structure of the three first highest singularities in 
$\epsilon$:
\begin{equation}
C^0(\epsilon)=C(\epsilon)Z(\epsilon).\label{epspoles}
\end{equation}
The key point of Ref.\cite{Vogt:2011jv} is that they realized that for example for the case 
of the gluon the small $\omega$ behavior of the bare coefficient function is
\begin{equation}
C^0(\epsilon)=\frac{1}{\omega}\,\sum_n \frac{\alpha_s^n}{\epsilon^{2n-1}}\sum_{l=0}^{n-1}\frac{1}{1-2(n-l)\epsilon/\omega}(A_{ln}+
\epsilon B_{ln}+\epsilon^2 C_{ln}+\dots).\label{vogtguess}
\end{equation}
Finally comparing Eq.(\ref{vogtguess}) with Eq.(\ref{epspoles}) one obtaines sytems of equations for 
the coefficients $A,B$ and $C$, which produce sequences up to arbitrary orders in $\alpha_s$
of the three highest powers in $1/\omega$ or equivalently (back to $x$-space) in $\ln x$. 
Then the highly non trivial
part of this approach is the solution of these sequences that are obtained for the large logarithms
and this has been successfully obtained in Refs.\cite{Vogt:2011jv,Kom:2012hd}.

\section{Plus-minus component evolution for the gluon-singlet system}

In Ref.\cite{Bolzoni:2012ii} it has been shown that, by use of the results obtained in Ref.\cite{Kom:2012hd}, the
scale dependence on $Q^2$ of the gluon and quark multiplicities defined by Eq.(\ref{multdef}) and governed by Eq.(\ref{ap})
can be computed analytically in a closed form. The approach that we used is a generalization of the techniques used in
Ref.\cite{Buras:1979yt} used to solve the evolution equation of the parton densities in the ``spacelike'' case. 
Our result can be written in the following way
\begin{eqnarray}
\langle n_h(Q^2)\rangle_{g}&=&\langle n_h(Q_0^2)\rangle_{g}\hat{T}_+(Q^2,Q_0^2),\nonumber\\
&&\nonumber\\
\langle n_h(Q^2)\rangle_{s}&=&\langle n_h(Q_0^2)\rangle_{g}\frac{\hat{T}_+(Q^2,Q_0^2)}{r_+(Q^2)}+
\left[\langle n_h(Q_0^2)\rangle_{s}-\frac{\langle n_h(Q_0^2)\rangle_{g}}{r_+(Q_0^2)}\right]
\hat{T}_-(Q^2,Q_0^2),\label{result}
\end{eqnarray}	 
where $Q_0^2$ is an arbitrary reference scale (see below). The dependence on $Q^2$ is always inside the running coupling constant 
$\alpha_s(Q^2)$. Indeed substituting the QCD values for the color factors and choosing $n_f=5$ in the formulae 
given in Ref.\cite{Bolzoni:2012ii} we can write that at NNLL
\begin{eqnarray}
\hat{T}_-(Q^2,Q_0^2)&=&\left[\frac{\alpha_s(Q^2)}{\alpha_s(Q_0^2)}\right]^{d_1} \nonumber\\
&&\,\nonumber\\
\hat{T}_+(Q^2,Q_0^2)&=&\exp\left\{d_2\left(\frac{1}{\sqrt{\alpha_s(Q^2)}}-\frac{1}{\sqrt{\alpha_s(Q_0^2)}}\right)
+d_3\left(\sqrt{\alpha_s(Q^2)}-\sqrt{\alpha_s(Q_0^2)}\right)\right\}\nonumber\\
&&\times\left[\frac{\alpha_s(Q^2)}{\alpha_s(Q_0^2)}\right]^{d_4},\label{exp}
\end{eqnarray}
where
\begin{eqnarray}
d_1=0.38647,\quad d_2=2.65187, \quad d_3=-3.87674,\quad d_4=0.97771. 
\end{eqnarray}
The function $r_{+}(Q^2)$ in Eq.(\ref{result}) has been computed in \cite{Dremin:1999ji,Capella:1999ms,Dremin:2000ep}
up to the third order (NNNLO) in the $\sqrt{\alpha_s}$ expansion and in QCD is given by:
\begin{equation}
r_{+}(Q^2)=2.25-0.61567\,\sqrt{\alpha_s(Q^2)}-2.19156\,\alpha_s(Q^2)+0.24348\,\alpha_s^{3/2}(Q^2)+\emph{O}(\alpha_s^2),\label{dremin}
\end{equation}
again with $n_f=5$. The first term in Eq.(\ref{dremin}) corresponds to the lowest order expectation for the ratio between the
gluon and the quark multiplicities given by $C_A/C_F$. According to Eq.(\ref{result}), our prediction for the multiplicity ratio $r$
is 
\begin{equation}
r(Q^2)=\frac{r_{+}(Q^2)}{1+\frac{r_{+}(Q^2)}{r_{+}(Q_0^2)}\left(\frac{\langle n_h(Q_0^2)\rangle_{s}r_{+}(Q_0^2)}{\langle n_h(Q_0^2)\rangle_{g}}
-1\right)\frac{\hat{T}_-(Q^2,Q_0^2)}{\hat{T}_+(Q^2,Q_0^2)}}.\label{result2}
\end{equation}
From this expression we can see that our result is a generalization of the result in Eq.(\ref{dremin}) due to the 
inclusion of the contribution proportional to
$\hat{T}_-(Q^2,Q_0^2)$ in the denominator of Eq.(\ref{result2}).

Now we show that our results in Eqs.(\ref{result},\ref{result2}) do not depend on the reference scale $Q_0^2$ and doing this also the
meaning of $\hat{T}_\pm(Q^2,Q_0^2)$ will become clear. We define the so called ``plus'' and ``minus'' component in terms of the singlet and
the gluon components defined in Eq.(\ref{multdef}) in the following way:
\begin{eqnarray}
\langle n_h(Q^2)\rangle_{+}&\equiv&\langle n_h(Q^2)\rangle_{g},\nonumber\\
\langle n_h(Q^2)\rangle_{-}&\equiv&\langle n_h(Q^2)\rangle_{s}-\frac{\langle n_h(Q^2)\rangle_{g}}{r_+(Q^2)}.
\end{eqnarray}
With this definition we have that the system of Eqs.(\ref{result}) is equivalent to the following one
\begin{equation}
\langle n_h(Q^2)\rangle_{\pm}=\langle n_h(Q_0^2)\rangle_{\pm}\hat{T}_\pm(Q^2,Q_0^2).\label{belle}
\end{equation}
By noting that according to Eqs.(\ref{exp}),
\begin{equation}
\hat{T}_\pm(Q^2,Q_0^2)\hat{T}_\pm(Q_0^2,Q_1^2)=\hat{T}_\pm(Q^2,Q_1^2),
\end{equation}
one can directly check that, after expressing $\langle n_h(Q_0^2)\rangle_{\pm}$ in terms of $Q_1^2$ using Eqs.(\ref{belle}) and then 
replacing them again into Eqs.(\ref{belle}), one arrives at the same expression but with $Q_0^2$ replaced by $Q_1^2$.
This shows the independence of the result on the reference scale $Q_0^2$ which is arbitrary and that $\hat{T}_\pm(Q^2,Q_0^2)$
just represent the renormalization group exponents in a basis where the splitting function matrix of Eq.(\ref{ap}) can be considered
diagonal. To look for a basis where the splitting function matrix could be considered diagonal
by a sufficient accuracy was the main motivation under the computation in Ref.\cite{Bolzoni:2012ii}. 

\section{Global fit to the avarage multiplicity data: determination of $\alpha_s$}

Our result Eqs.(\ref{result}) depends on three quantities that should be fixed by the comparison with the experiment.
They are $\langle n_h(Q_0^2)\rangle_{g}$, $\langle n_h(Q_0^2)\rangle_{s}$ (representing the avarage multiplicity
at the reference scale $Q_0^2$) 
and a reference value for the running coupling constant
which is conventionally chosen at the scale of the mass of the $Z^0$ weak boson, $\alpha_s(M_Z)$. In Ref.\cite{Bolzoni:2012ii},
$\alpha_s(M_Z)$ has been fixed as an input parameter to be equal to the world avarage value \cite{Bethke:2012jm}. 
The result that we obtained was 
\begin{eqnarray}
\langle n_h(Q_0^2)\rangle_{g}&=&24.02\pm0.36,\quad 
\langle n_h(Q_0^2)\rangle_{g}=15.83\pm0.37,\quad\rm{ 90\%\,\,\,C.L.},\nonumber\\
\alpha_s(M_Z)&=&0.118,\quad\rm{input\,\, parameter},
\label{fitnnll}
\end{eqnarray}
where the arbitrary scale has been chosen to be $Q_0=50\,\, \rm{GeV}$. We have checked that the $\chi^2$ remains unchanged 
changing the value of $Q_0^2$ as expected from the considerations done above.

Due to the quite 
good agreement with the data obtained in Ref.\cite{Bolzoni:2012ii}, we tried to treat $\alpha_s(M_Z)$ as an additional fit parameter.
The result that we obtained for such a fit (always for $Q_0=50\,\, \rm{GeV}$) is:
\begin{eqnarray}
\langle n_h(Q_0^2)\rangle_{g}&=&24.18\pm0.32, \quad
\langle n_h(Q_0^2)\rangle_{g}=15.86\pm0.37,\quad\rm{ 90\%\,\,\,C.L.},\nonumber\\
\alpha_s(M_Z)&=&0.124\pm 0.005,\quad\rm{ 90\%\,\,\,C.L.},
\label{fitnnllas}
\end{eqnarray}
which is in agreement with Eq.(\ref{fitnnll}) and with the experimental values for the gluon and quark multiplicities 
at $50\,\, \rm{GeV}$.
The $\chi^2$ per degree of freedom is improved in this fit from $3.71$ (corresponding to Eqs.(\ref{fitnnll})) 
down to $2.84$ (corresponding to Eqs.(\ref{fitnnllas})). 
The predicted value for $\alpha_s(M_Z)$ is higher than the world avarage value \cite{Bethke:2012jm}, however such
a higher value is not new in hadronic final states of electron-positron annihilation \cite{Dissertori:2009ik}.
However how much it would make sense to compare our result with $\alpha_s^{\overline{MS}}$ is still under investigation.
For the determination of the result in Eqs.\ref{fitnnllas} we have used the same data selection we used in 
Ref.\cite{Bolzoni:2012ii}, which is also the one adopted by the DELPHI colaboration in Ref.\cite{Abdallah:2005cy}. 
\begin{figure}
\centering
\includegraphics[scale=0.7]{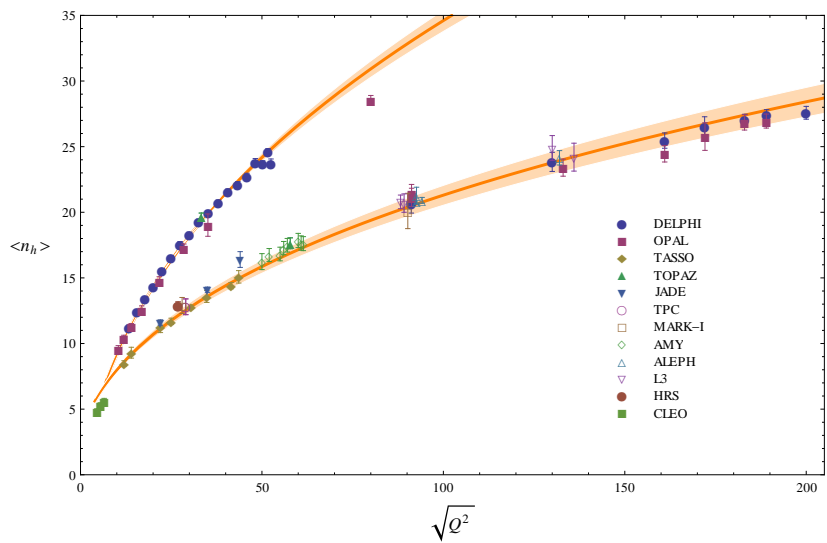}
\caption{Gluon and quark multiplicities fits according to Eq.(4.2) compared to the data.}
\label{Fig:as_mult}
\end{figure}
In Fig.\ref{Fig:as_mult} we plot the result of our new fir according to Eq.(\ref{fitnnllas})
together with the uncertainties and the data, which are taken from the Tables in Ref.\cite{Siebel:2003zz}.

\section{Conclusions} 
In this talk we have sketched the history and the various techniques culminating in the very important new results
from the theoretical side (see Refs.\cite{Almasy:2011eq,Kom:2012hd} and referencies therein for the most 
recent developments),
which have already made possible new important improvement in the descriprion of the data in hadronic final states processes
in the ``timelike'' case (see Ref.\cite{Bolzoni:2012ii}).
Motivited by the well description of the data obtained in Ref.\cite{Bolzoni:2012ii} for the  avarage 
multiplicities in jets initiated by gluon and quarks, we have extended our previous analysis making the value
of the strong coupling constant at the reference scale of the mass of the $Z^0$ boson a parameter for the fit. 
In this way we have still obtained numbers in agreement with observations 
for the value of the multiplicities at the reference scale $Q_0=50\,\, \rm{GeV}$ and in addition we got
a significant improvement in the $\chi^2$ per degree of freedom and, as a byproduct, a prediction for $\alpha_s(M_Z)$.
\\
\\
{\bf{Acknowledgements}}:
I would like to thank the organizers of the ``XXI International Baldin Seminar on High Energy Physics Problems''
and A. Kotikov for fruitful discussions and for inviting me to this conference. This work is in part 
supported by the Heisenberg-Landau 
program and in part by the German Federal Ministry for Education
and Research BMBF through Grant No.\ 05~HT6GUA, by the German Research
Foundation DFG through the Collaborative Research Centre No.~676
{\it Particles, Strings and the Early Universe---The Structure of Matter and
Space Time}, and by the Helmholtz Association HGF through the Helmholtz
Alliance Ha~101 {\it Physics at the Terascale}.


\end{document}